\documentclass[aps,prb,twocolumn,superscriptaddress]{revtex4-2}
\usepackage{amsmath, amssymb,  graphicx,bbm}

\usepackage[colorlinks=true ,urlcolor=blue,urlbordercolor={0 1 1}]{hyperref}

\usepackage{color}
\usepackage{xcolor}
\usepackage{braket}

\usepackage[utf8]{inputenc}
\usepackage{bm}
\usepackage{verbatim}
\usepackage{graphicx}
\usepackage{amsmath}
\usepackage{color}
\usepackage{float}
\usepackage{bbm}
\usepackage{amssymb}
\usepackage{slashed}
\usepackage{wasysym,bm,bbm,dsfont,braket}

\begin{document}
\title{Magnetism and superconductivity in bilayer nickelate}
\author{Hui Yang}
\affiliation{Department of Physics and Astronomy, Johns Hopkins University, Baltimore, Maryland 21218, USA}
\affiliation{Department of Physics and Astronomy, University of Pittsburgh, PA 15213, USA}
\author{Ya-Hui Zhang}
\affiliation{Department of Physics and Astronomy, Johns Hopkins University, Baltimore, Maryland 21218, USA}
\begin{abstract}
The discovery of high-temperature superconductivity in bilayer nickelate La$_{3}$Ni$_{2}$O$_{7}$ necessitates a minimal theoretical model that unifies the  superconducting phase with the  spin-density-wave (SDW) phase without external pressure or strain. We propose a model where half-filled $d_{z^{2}}$ local moments interact with itinerant $d_{x^{2}-y^{2}}$ electrons via strong Hund's coupling $J_H$, which reduces to a bilayer type-II t-J model in the large $J_H$ limit. Using iDMRG calculations on an $L_y=4, L_z=2$ cylinder, we demonstrate that the competition between double-exchange ferromagnetism and in-plane superexchange generates period-4 stripe-like SDW order—a feature absent in  one-orbital t-J model with only $d_{x^2-y^2}$ orbital. Furthermore, increasing the interlayer exchange coupling suppresses magnetic order and stabilizes interlayer s-wave superconductivity. These results identify the type-II t-J model as a minimal framework for capturing the interplay of magnetism and superconductivity in bilayer nickelates.
\end{abstract}
\maketitle

\textbf{Introduction} High-temperature superconductivity has been a central topic in condensed matter physics for decades~\cite{lee2006doping}. Recently, a new family of high-temperature superconductors with $T_c\approx 80K$~\cite{sun2023signatures} was discovered in the bilayer nickelate La$_3$Ni$_2$O$_7$ under high pressure. Remarkably, superconductivity can also emerge at ambient pressure in epitaxially strained thin films~\cite{Ko2025,Zhou2025}. In addition to superconductivity, a variety of correlated phenomena have been observed in this system, including linear-in-temperature resistivity~\cite{sun2023signatures,Zhang2024_str,Hou_2023}, magnetic order~\cite{Chen2024,gupta2024anisotropicspinstripedomains,Ren2025,PhysRevLett.132.256503,Khasanov2025,FUKAMACHI2001195,doi:10.7566/JPSJ.93.053702,Zhao_2025,plokhikh2025unravelingspindensitywave}, and a rich normal-state phase diagram. These discoveries have stimulated intense interest in understanding the underlying electronic structure and interaction mechanisms in bilayer nickelates. Unlike cuprates, bilayer nickelates exhibit several distinctive features: a strongly coupled bilayer structure, substantial interlayer hybridization, and an orbital-selective electronic configuration dominated by the $d_{x^2-y^2}$ and $d_{z^2}$ orbitals. These characteristics suggest that interlayer interactions and multiorbital physics play a crucial role in determining the ground-state properties. Considerable experimental efforts~\cite{sun2023signatures,Zhang2024_str,Hou_2023,PhysRevX.14.011040,wang2023observation,ZHANG2024147,zhou2024investigationskeyissuesreproducibility,Wang2024,Wang2024_str,10.1093/nsr/nwaf220,PhysRevLett.133.146002,Dong2024,Chen2024,Chen2024_chemi,Wang2025,Li_2024,li2024distinguishing,zhou2024revealing,Yang2024,Khasanov2025,Wang2025,Ko2025,Zhou2025,Huo2025,Liu2025,10.1093/nsr/nwaf205,10.1093/nsr/nwaf253,bhatt2025resolving,wang2025electronic,sun2025observation,li2025enhanced,Hao2025,fan2025superconducting,shen2025anomalous,wang2025electron}
and theoretical studies~\cite{oh2023type,lu2023interlayer,qu2024bilayer,yang2024strong,lange2023pairing,luo2023bilayer,zhang2023electronic,huang2023impurity,Zhang2024,Geisler2024,PhysRevMaterials.8.044801,PhysRevB.109.045151,sakakibara2023possible,tian2024correlation,qin2023high,yang2023minimal,zhan2024cooperation,chen2024non,yang2023possible,gu2023effective,liu2023s,shen2023effective,PhysRevB.109.104508,PhysRevB.109.205156,PhysRevB.109.L201124,oh2023type,zhang2023strong,zhu2025quantum,pan2023effect,PhysRevB.110.024514,PhysRevB.110.L060510,PhysRevB.110.104507,PhysRevB.110.094509,Luo2024,PhysRevB.109.045127,PhysRevB.109.045154,PhysRevLett.133.096002,Ouyang2024,PhysRevB.108.125105,lange2023pairing,cao2023flat,qu2023bilayer,PhysRevB.108.214522,zhang2023trends,PhysRevB.111.014515,PhysRevB.109.115114,PhysRevB.110.205122,PhysRevB.109.L180502,tian2025spin,liu2025origin,liao2024orbital,PhysRevLett.132.126503,yin2025s,PhysRevB.111.104505,kaneko2025t,ji2025strong,Wang_2025,haque2025dft,shi2025theoretical,gao2025robust,le2025landscape,hu2025electronic,shao2024possible,rm9g-8lm1,ushio2025theoretical,duan2025orbital,qiu2025pairing,cao2025strain,shao2025pairing,PhysRevB.111.L020504,xue2024magnetism,qu2024bilayer,lu2023superconductivity,lu2023interlayer,Fan2025_nickelate}
have been devoted to dissecting the magnetism and superconductivity in this material.  In particular, neutron and 
$\mu$SR experiments have observed a spin-density-wave (SDW) with ordering vector ${\bf Q}=(\pi/2,\pi/2)$ at ambient pressure without strain~\cite{Chen2024,gupta2024anisotropicspinstripedomains,Ren2025,PhysRevLett.132.256503,Khasanov2025,FUKAMACHI2001195,doi:10.7566/JPSJ.93.053702,Zhao_2025,plokhikh2025unravelingspindensitywave}. 
Despite these advances, the evolution from the SDW order to superconductivity remains incompletely understood.

A variety of theoretical proposals have been put forward to explain the pairing mechanism in bilayer nickelates. It is clear that the active degrees of freedom are the two e$_g$ orbitals: $d_{x^2-y^2}$ orbital and $d_{z^2}$ orbital of the nickel atom. In one scenario~\cite{oh2023type,lu2023interlayer,lu2023superconductivity,qu2024bilayer,yang2024strong,zhang2023strong,lange2023pairing}, the interlayer spin-coupling $J_\perp$ of $d_{z^2}$ orbital is shared to the $d_{x^2-y^2}$ orbital through the large on-site Hund's coupling, which leads to interlayer s-wave pairing of the $d_{x^2-y^2}$ orbital. However, it is not clear whether a one-orbital model with only the $d_{x^2-y^2}$ orbital is sufficient. In this work, we argue that the local moments from the $d_{z^2}$ orbital are necessary to explain the experimentally observed magnetic order at ambient pressure and thus cannot be ignored in theoretical studies.

\begin{figure}[h]
    \centering
\includegraphics[width=1.0\linewidth]{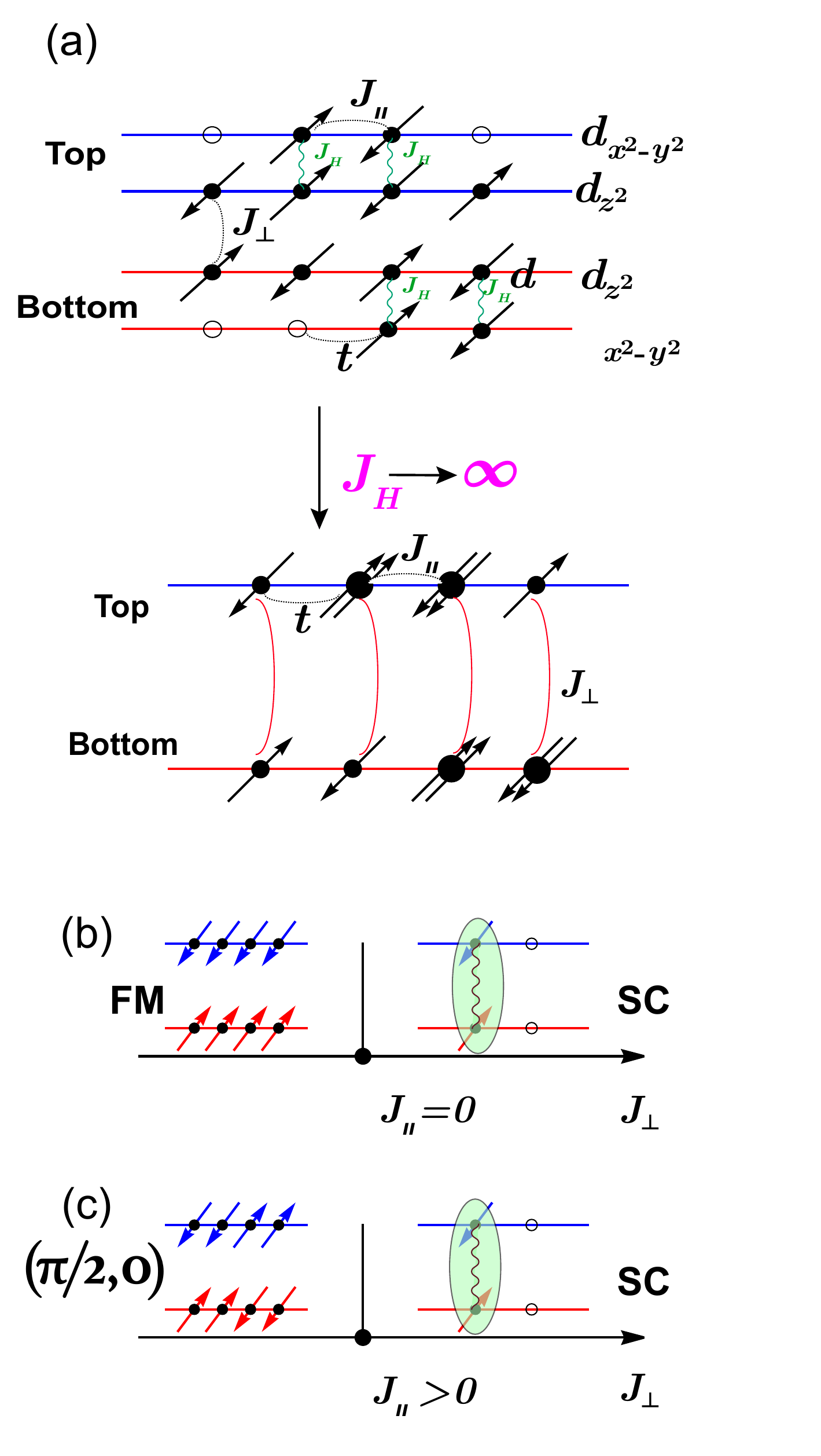}
\caption{(a) Illustration of the double Kondo model and type-II t-J model in the limit $J_H\rightarrow\infty$. The single arrow and double arrow correspond to the spin-$1/2$ and spin-$1$ moment, respectively. (b), (c) schematic phase diagrams of the bilayer type-II t-J model for $J_\parallel=0$ (b) and finite $J_\parallel>0$ (c). At small $J_\perp$, there is a magnetic ordered phase. As the interlayer coupling $J_\perp$ increases, the magnetism is suppressed and an interlayer s-wave superconductor emerges. }
\label{fig:phase_diagram}
\end{figure}

The central question is whether one can construct a minimal microscopic model that simultaneously captures the SDW phase observed at ambient pressure and the superconducting phase that emerges under pressure or strain? Conventional two-orbital models are computationally prohibitive for large-scale numerical simulations due to their expansive Hilbert spaces. To address this, we propose the bilayer type-II $t$-$J$ model~\cite{zhang2020type,oh2023type} as a minimal alternative. Derived in the large $J_H$ limit (see Fig.~\ref{fig:phase_diagram}(a)), this model retains only five states per site: two spin-half Ni$^{3+}$ and three spin-one Ni$^{2+}$ states. Using infinite DMRG (iDMRG), we uncover a magnetically ordered phase at weak interlayer coupling—a feature notably absent in the one-orbital bilayer $t$-$J$ model—as well as an interlayer superconducting phase stabilized by stronger interlayer exchange. As illustrated in the phase diagram (Fig.~\ref{fig:phase_diagram}(b) and (c)), when the interlayer coupling $J_\perp$ is small, the ground state is either a ferromagnetic phase or a period-4 SDW phase, depending on the intralayer coupling $J_\parallel$. Increasing $J_\perp$ suppresses magnetic order and facilitates the emergence of interlayer $s$-wave superconductivity. We propose that the primary experimental role of pressure or strain is to enhance interlayer coupling by straightening the buckled vertical Ni-O bonds. Collectively, our results provide a unified description of magnetism and superconductivity in bilayer nickelates and establish the bilayer type-II $t$-$J$ model as a minimal framework for future theoretical investigations.

\textbf{Model}  As illustrated in Fig.~\ref{fig:phase_diagram}(a), there are two active $3d$-orbitals in La$_3$Ni$_2$O$_7$: the itinerant $d_{x^2-y^2}$ band and the more localized $d_{z^2}$ band. Because the $d_{z^2}$ orbital has lower energy, it is nearly half-filled and forms a Mott insulator, contributing a spin-$1/2$ moment on each Ni site. Its large interlayer hopping generates a sizable interlayer super-exchange $J_\perp$. In contrast, the $d_{x^2-y^2}$ orbital hosts the doped carriers moving within each Ni-O plane. These itinerant electrons are strongly coupled to the localized $d_{z^2}$ moments by the Hund’s coupling $J_H$. This dual nature—localized $d_{z^2}$ spins interacting with itinerant $d_{x^2-y^2}$ carriers, can be described by a bilayer Kondo lattice model~\cite{yang2024strong,PhysRevB.111.L241102,oh2025hightemperaturesuperconductivitykineticenergy} (see Fig.~\ref{fig:phase_diagram}(a)).

The valence of Ni is at $3d^{8-x}$ with $x\approx 0.5$, thus the system is in a superposition of $d^8$ and $d^7$ configurations. We are interested in the large $J_H$ limit.  In this case, $d^{8}$ configuration hosts a spin-1 moment formed by one electron in the $d_{x^2-y^2}$ orbital together with the local moment in the $d_{z^2}$ orbital. On the other hand, the $d^7$  configuration has just the localized spin-1/2 moment in the $d_{z^2}$ orbital. In total, we have a  $5$-dimensional local Hilbert space at each site per layer in the large $J_H$ region. Projecting the bilayer Kondo model onto this restricted Hilbert space, we arrive at the bilayer type-II $t-J$ model~\cite{zhang2020type,PhysRevB.111.L020504},
    \begin{align}
     H=
     -&t\sum_{l,\sigma,\langle ij \rangle}
\left[Pc^{\dagger}_{i;l;\sigma}
c_{j;l;\sigma}P
+h.c.\right]
     \notag\\
     +&  \sum_{l,\langle ij \rangle}
   \left[
   J^{ss}_\parallel
\vec S_{i;l}\cdot \vec S_{j;l}
      +
      J^{dd}_\parallel
\vec T_{i;l}\cdot \vec T_{j;l} \notag 
      \right]
      \notag\\
+& \sum_{l,\langle ij \rangle}J^{sd}_\parallel
      \left[\vec S_{i;l}\cdot \vec T_{j;l} +\vec T_{i;l}\cdot \vec S_{j;l}\right]
      \notag\\
+& \sum_{i} \left[J^{ss}_\perp\vec S_{i;t}\cdot \vec S_{i;b}
+J ^{dd}_\perp  \vec T_{i;t}\cdot \vec T_{i;b} \right]\notag\\
+& \sum_{i} J^{sd}_\perp
\left[\vec S_{i;t}\cdot \vec T_{i;b}+\vec T_{i;t}\cdot \vec S_{i;b}\right],
\label{eq:type_II_t_J}
\end{align}
where $c^\dagger_{i;l;\sigma}$ is the projected electron operator of the $d_{x^2-y^2}$ orbital. $l=t,b$ labels the layer and $\sigma=\uparrow,\downarrow$ labels the spin. $\vec{S}$ and $\vec{T}$ are the spin operators for the singlon ($d^7$) and doublon ($d^8$) Hilbert space with $\vec{S}=\frac{1}{2}\ket{\sigma}\vec{\sigma}_{\sigma\sigma^\prime}\bra{\sigma^\prime}$, $\vec{T}=\ket{a}\vec{T}_{ab}\bra{b}$, respectively, here $a,b=1,0,-1$ is the $S_z$ component of the spin-$1$ moment. $\vec T$ can be found in the supplementary.  The interlayer coupling is from the $J_\perp$ term of $d_{z^2}$ orbital, and we have $J_\perp^{ss}=2J_\perp^{sd}=4J_\perp^{dd}=J_\perp$. In contrast, the intralayer spin coupling is from the $d_{x^2-y^2}$ orbital and we have $J_\parallel^{ss}=J_\parallel^{sd}=0$ and $J_\parallel^{dd}=\frac{1}{4}J_\parallel$, where $J_\parallel$ is the super-exchange for the $d_{x^2-y^2}$ orbital.

In the following we are going to study this model using iDMRG~\cite{PhysRevLett.69.2863,tenpy}. We use two conservation numbers in the DMRG simulation: the $U(1)$ charge conservation and $U(1)$ $S_z$ conservation. The correlation length can be calculated from the transfer matrix method in different symmetry sectors. In our DMRG simulation, the system size is up to $L_y=4$ and $L_z=2$, and the bond dimension is up to $m=10000$ with the truncation error smaller than $10^{-4}$.

\begin{figure}
    \centering
\includegraphics[width=1.0\linewidth]{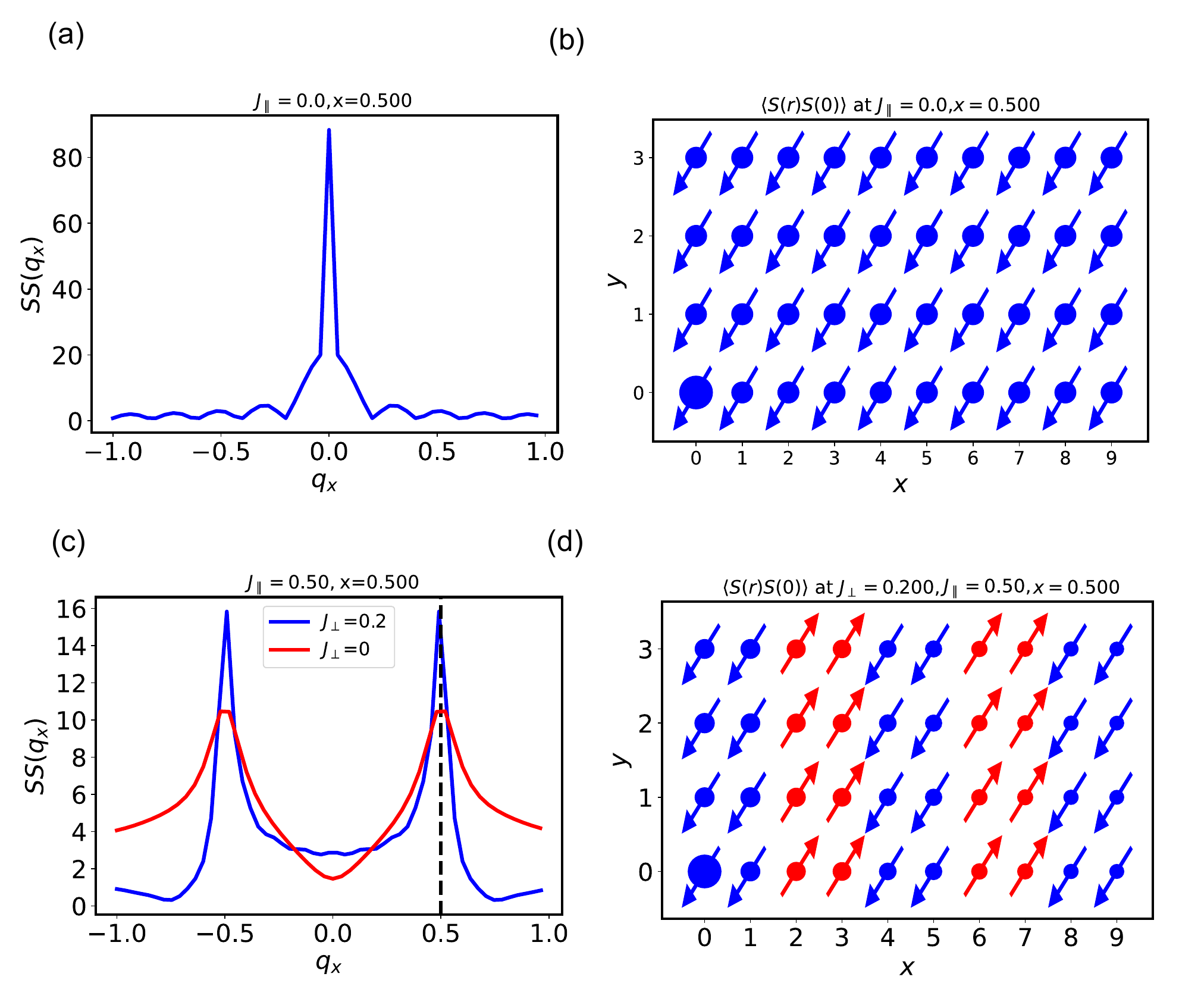}
\caption{iDMRG results  for type II  t-J model with $L_y=4, L_z=2$ and hole doping $x=0.5$. (a) spin structure factor $\langle\vec{S}(q_x,q_y=0)\cdot \vec{S}(-q_x,q_y=0)\rangle$ for single layer model with $J_\parallel=0$; (b) real space spin-spin correlation function for single layer type-II t-J model with $J_\parallel=0$; (c) the spin-spin correlation function $\langle \vec{S}(q_x,q_y=0)\cdot\vec{S}(-q_x,q_y=0)\rangle$ with $J_{\parallel}=0.5$ for single layer (red line) and bilayer (blue line) type-II t-J model with $J_\perp=0.2$; (d) the spin-spin correlation in real space for $J_{\parallel}=0.5, J_{\perp}=0.2$.  In real-space spin-spin correlation plot,  the size of the circle indicates the value of the correlation function $\langle \vec S(\mathbf r)\cdot \vec S(0)\rangle$, and the arrow indicates the direction of the spin configuration. $q_x$ is in units of $\pi$. In the plot of (c), we scale the data for single layer correlation (red line) to $4\langle\vec{S}({\bf q})\cdot\vec{S}(-{\bf q})\rangle$. The bond dimension is $m=7000$ for single layer model and $m=10000$ for bilayer model.}
\label{fig:spin_spin_Ly=4}
\end{figure}

\textbf{Magnetism at $J_\perp=0$ limit}  We start from the decoupling limit with $J_\perp=0$. Effectively we can consider a single-layer Kondo lattice model with Kondo coupling $J_K=-2J_H \rightarrow -\infty$. It is known that the ground state is in a spin-polarized phase due to the double-exchange mechanism~\cite{PhysRev.118.141}  when $J_{\parallel}=0$. The DMRG results in this limit are summarized in Fig.~\ref{fig:spin_spin_Ly=4}(a) and (b):  the spin structure factor $\langle\vec{S}({\bf q})\cdot\vec{S}(-{\bf q})\rangle$ exhibits a pronounced peak at $\bf{q}=(0,0)$ and the real space correlations confirm the robust ferromagnetic order.

However, the existence of the finite $J_\parallel$ terms favors the Neel order with $\mathbf{Q}=(\pi,\pi)$, which competes with the kinetic energy driven ferromagnetism with $\mathbf Q=(0,0)$. As a result, the system makes a compromise and is in an SDW order with a long wavelength. As shown in Fig.~\ref{fig:spin_spin_Ly=4}(c) and (d), a finite $J_\parallel$  drives the system into a stripe order with period four. In particular the peak of spin structure factor  shifts from ${\bf Q}=(0,0)$ to   ${\bf Q}=(\pi/2,0)$. Both the FM order and  this stripe order arise from the local moments of the $d_{z^2}$ orbital. Under a small $J_\perp$,  each layer remains in the SDW order, as shown in Fig.~\ref{fig:spin_spin_Ly=4}(d) for $J_\perp=0.2 \ t$.

\textbf{Interlayer pairing at $L_y=2$}  In the following we show that there is a phase transition from the magnetic order at $J_\perp=0$ limit discussed above to a superconductor at large $J_\perp$. Strong pairing has been found in DMRG simulation of the model at $L_y=1$. But here we will show the DMRG evidences of pairing for $L_y=2$ and $L_y=4$ cylinders.

 We first perform finite DMRG calculations on a system with dimensions $L_x=20$, $L_y=2$, and $L_z=2$. We compute the spin gap, defined as $\Delta_S = E(S^z = 1) - E(S^z = 0)$. The results for doping $x = 0.1$ are summarized in Table~\ref{tabel:spin_gap}. For small interlayer coupling $J_\perp \le 0.1$, we find $\Delta_S \approx 0$ in type-II t-J model, signaling the presence of gapless spin excitations, and our iDMRG at $x=0.1$ yields a central charge around $c\approx 3$ as shown in the supplementary, indicating this phase is a multi-component Luttinger liquid phase.  In contrast, for $J_\perp \ge 0.3$, a finite spin gap develops, signaling the emergence of a Luther–Emery liquid phase~\cite{PhysRevLett.33.589}—the one-dimensional analogue of a superconducting state.   As a comparison, we also show the spin gap of one-orbital t-J model with only $d_{x^2-y^2}$ orbital (see the supplementary)  in Table.~\ref{tabel:spin_gap}. At large $J_\perp$ the spin gap in the one-orbital model with the same $J_\perp$ is $4-6$ times larger than that in the type-II t–J model, indicating that the one-orbital model substantially overestimates the superconducting tendency. At small $J_\perp$, the discrepancy becomes even more pronounced: the one-orbital t–J model exhibits a finite spin gap—signaling intralayer pairing—while the type-II t–J model remains gapless in the spin sector.  This contrast highlights that the type-II t–J model, which keeps the localized $d_{z^2}$ moments, provides a more realistic minimal model for describing the behavior of bilayer nickelates in both the small $J_\perp$ and large $J_\perp$ regime.

\begin{table}[h]
\centering
\begin{tabular}{ c|c c c c c c}
\hline
\hline
  $J_\perp/t$&$0.1t$ & $0.3t$ & $0.5t$ & $1.0t$ & $2.0t$ \\ 
  \hline
 type-II  & $0.00414t$ & $0.02067t$ & $0.0455t$ & $0.08921t$ & $0.22854t$\\
 \hline
 one-orbital  & $0.12217t$ & $0.09222t$ & $0.18825t$  & $0.49628t$ & $1.31314t$\\
 \hline
 \hline
\end{tabular}
\caption{The spin gap $\Delta_S=E(S^z=1)-E(S^z=0)$ for various $J_\perp/t$ with $L_x=20$, $L_y=2$, $L_z=2$ at $x=0.1$ for bilayer type II t-J model and bilayer one-orbital t-J model. We use $J_\parallel=0.5 t$ for both models. 
The results are achieved by finite DMRG simulation with bond dimension  $m=7000$ for $J_\perp > 0.5$, $m=10000$ for $J_\perp=0.1$ and $m=15000$ for $J_\perp=0.3-0.4$.  In large $J_\perp$, both models exhibit interlayer pairing, but the one-orbital model overestimates the pairing gap by a factor of $4-6$. For small $J_\perp<0.3$, the two models are qualitatively different. The one-orbital model shows a pairing gap from intralayer pairing, while the type II t-J model is in a Luttinger liquid phase.  } 

\label{tabel:spin_gap}
\end{table}

\begin{figure}[ht]
    \centering
\includegraphics[width=1.0\linewidth]{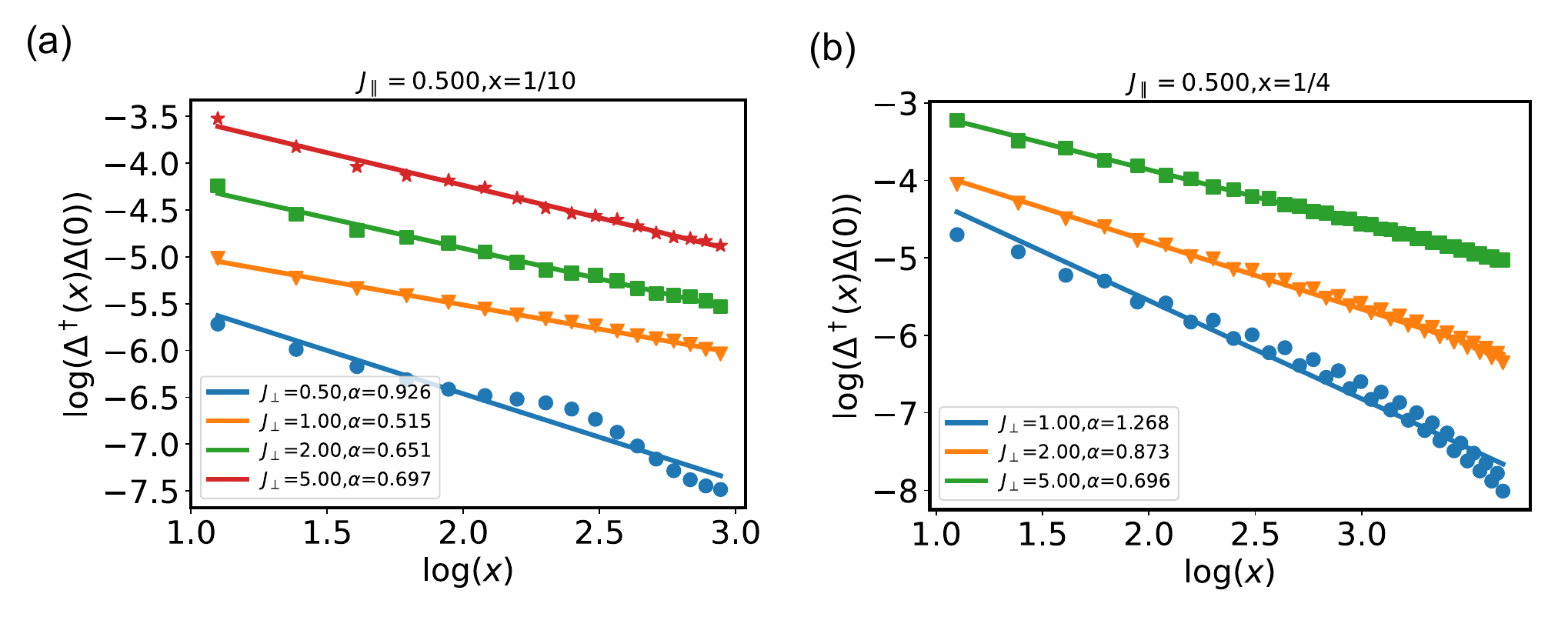}
\caption{The log-log plot of pair-pair correlation function for $L_y=2$, with $t=1$, $J_\parallel=0.5$ and hole doping (a) $x=0.1$, (b) $x=0.25$. The bond dimension is $m=10000$.}
\label{fig:sc_Ly=2}
\end{figure}

 The interlayer Cooper pair is defined as $\Delta_y(x)=\epsilon_{\sigma\sigma^\prime}c_{y,t,\sigma}(x)c_{y,b,\sigma^\prime}(x)$, where $y$ is the coordinate in $y$-direction. The pair–pair correlation functions $\langle\Delta_{y=0}^\dagger(x)\Delta_{y=0}(0)\rangle$ obtained from iDMRG on cylinders with $L_y=2$ for hole dopings $x=0.1$ and $x=0.25$ are shown in Fig.~\ref{fig:sc_Ly=2}(a) and (b), respectively. For $x=0.1$, we find that at $J_\perp \geq 0.5$, the interlayer pairing exhibits a robust power-law decay with exponent $\alpha<1$, consistent with the finite spin gap obtained from finite DMRG (Table~\ref{tabel:spin_gap}). For $x=0.25$, however, a slightly larger $J_\perp\ge1$ is required to stabilize power-law superconducting correlations. At even higher doping $x=0.5$, the behavior changes qualitatively. As shown in the supplementary, the system becomes an insulator at large $J_\perp$, with one interlayer Cooper pair per unit cell.

\begin{figure}[ht]
\centering
\includegraphics[width=1.0\linewidth]{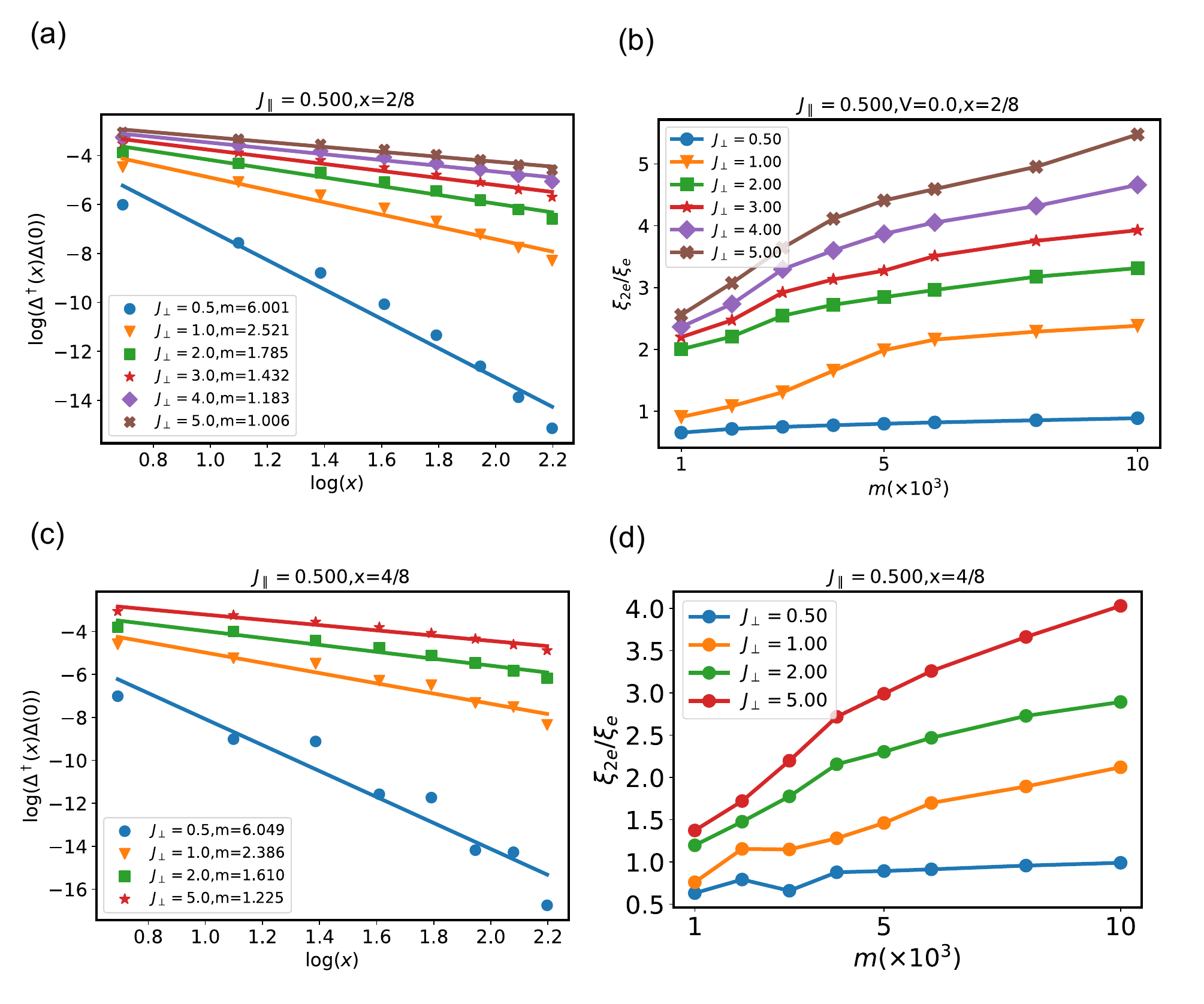}
\caption{ iDMRG results for bilayer type II t-J model with  $L_y=4$ and $J_\perp=0.5$, it is achieved with bond dimension $m=10000$. (a) the log-log plot of interlayer pair-pair correlation function for $L_y=4$, with $J_\parallel=0.5$ and hole doping $x=0.25$ at large $J_\perp$. (b) the ratio of correlation length $\xi_{2e}/\xi_e$ at $x=0.25$.  (c) and (d) same plot as (a) and (b) at doping $x=0.5$. The interlayer pairing operator is defined as $\Delta(\mathbf r)=\epsilon_{\sigma\sigma^\prime}c_{t,\sigma}(\mathbf r)c_{b,\sigma^\prime}(\mathbf r)$.}
\label{fig:sc_Ly=4}
\end{figure}

\textbf{Interlayer pairing at $L_y=4$} We then move to $L_y=4$. As shown in Fig.~\ref{fig:sc_Ly=4}, the interlayer pair–pair correlation function exhibits robust power-law decay when $J_\perp\geq 1$, indicating  interlayer paired s-wave superconductor at large $J_\perp$. At $x=0.5$, the dominant correlation length changes from spin to Cooper pair when we change $J_\perp$ from $0.5$ to $1$.   To characterize pairing, we also calculate the correlation length from iDMRG for pairing $\xi_{2e}$ and for single electron $\xi_e$. The growth of the ratio $\xi_{2e}/\xi_e$ with bond dimension is shown in Fig.~\ref{fig:sc_Ly=4}(b) and (d) for $x=0.25$ and $x=0.5$. For a Fermi liquid we expect the ratio $\frac{\xi_{2e}}{\xi_e}\approx 0.5$, which is indeed true for $J_\perp=0.5$.  In contrast, for $J_\perp\geq 1$, $\xi_{2e}/\xi_e$ grows with the bond dimension and is larger than $2$, indicating formation of Cooper pairs.  For both $x=0.25$ and $x=0.5$, the interlayer pairing correlation function show a power-law with exponent smaller than $2$ when $J_\perp\geq 2$ (see Fig.~\ref{fig:sc_Ly=4}(a)(c), indicating a divergent pairing susceptibility.   For $x=0.5$, we confirm that $J_\perp=0.5$ is still in the SDW phase. The SDW-SC transition occurs at $J_{\perp,c}\in (0.5,1)$, but a dense sampling of $J_\perp$ is numerically challenging to study the nature of the phase transition or the possibility of coexisting phase.

\begin{figure}[ht]
    \centering
\includegraphics[width=0.9\linewidth]{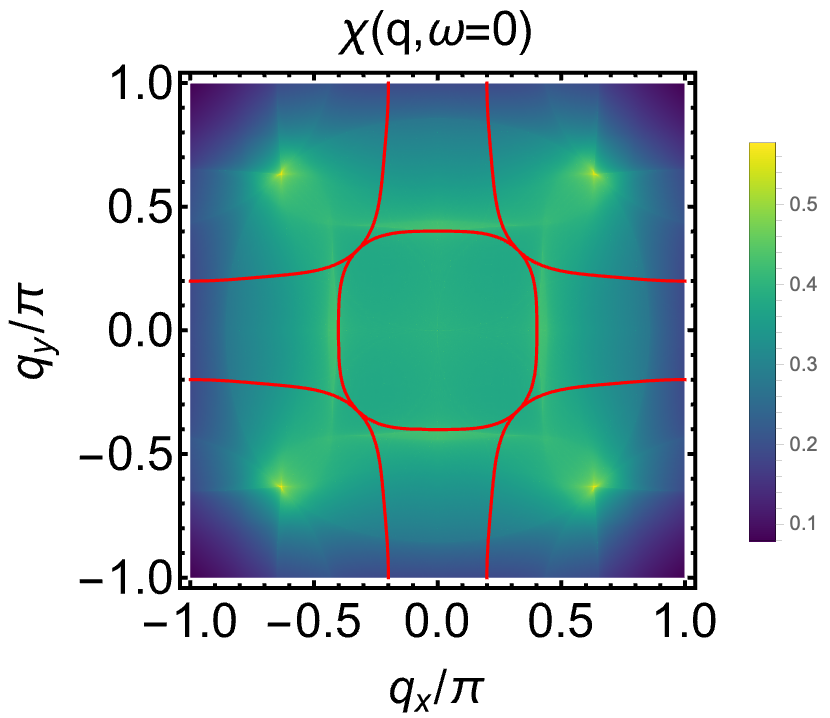}
\caption{The spin susceptibility $\chi^{tt}(q,\omega=0)$ from itinerant electrons. The peak is around ${\bf Q}=(0.632\pi,0.632\pi)$. The red line correspond to two Fermi surfaces from the dispersion. To fit the $\alpha$ and $\beta$ Fermi pockets, we add the interlayer hopping term as $t^\perp_{{\bf k}}=4t^\perp(\cos(k_x)-\cos(k_y)+\eta(\cos(2k_x)-\cos(2k_y)))^2$ and use the parameters from Ref.~\cite{PhysRevB.110.024514}. The nearest, next-nearest neighbor, second-next-nearest neighbor intralayer hoppings are $t=0.29eV$, $t_2/t=-0.18$, $t_3/t=0.05$,  and we use $t^\perp/t=-0.14$ and $\eta=0.15$.}
\label{fig:chi}
\end{figure}

\textbf{Discussion} Based on iDMRG simulations of the bilayer type-II $t$-$J$ model, we demonstrate the existence of spin-density wave (SDW) order at small $J_\perp\leq 0.5$ and a superconducting phase when $J_\perp \geq 1$ for $L_y=4$ cylinder at $x=0.5$.  Our results offer a plausible explanation of the experiment in bilayer nickelate. We propose that the primary role of pressure or strain is to enhance $J_\perp$ by straightening the buckled Ni-O bonds along the $z$-direction following the structural transition. Consequently, the experimental observation of SDW order at ambient conditions and a superconducting phase under pressure is consistent with the low- and high-$J_\perp$ regimes, respectively. A discrepancy remains regarding the stripe orientation: experiments observe stripes along the diagonal ($45^\circ$) direction, whereas our calculations yield an SDW along the $x$-direction. Theoretically, the SDW arises from a competition between double-exchange ferromagnetism and the Néel order induced by $J_{\parallel}$. We conjecture that the energy landscape with respect to the SDW momentum $\mathbf{Q}$ is relatively flat, making the exact ordering vector sensitive to microscopic details not included in our minimal model. Two specific perturbations could favor the experimentally observed $\mathbf{Q}=(\frac{\pi}{2},\frac{\pi}{2})$: (1) The system lacks $C_4$ rotational symmetry at ambient pressure; as noted in Ref.~\cite{wang2025originspinstripesbilayer}, this anisotropy may favor diagonal SDW ordering. (2) In realistic materials, an interlayer hopping term $\delta H=4t^\perp(\cos(k_x)-\cos(k_y)+\eta(\cos(2k_x)-\cos(2k_y)))^2 c^\dagger_{t;\sigma}(\mathbf{k})c_{b;\sigma}(\mathbf{k}) + \text{h.c.}$ splits the bilayer Fermi surfaces into $\alpha$ and $\beta$ pockets (see Fig.~\ref{fig:chi}). The magnetism of local moments from the $d_{z^2}$ orbital is influenced by the kinetic energy of the itinerant $d_{x^2-y^2}$ electrons. The itinerant electrons can mediate a RKKY interaction between the localized moments with $H_{RKKY}=-\sum_{\langle ij\rangle}J_{ij,ll^\prime}\vec{S}_{i;l}\cdot\vec{S}_{i;l^\prime}$, where $J_{ij,ll^\prime}$ is proportional to the susceptibility $\chi_S({\bf q})$ of the itinerant electrons. As shown in Fig.~\ref{fig:chi}, the maximum of the spin susceptibility is at momentum ${\bf Q}=(0.632\pi,0.632\pi)$, which is close to ${\bf Q}=(0.5\pi,0.5\pi)$ observed in experiments. Therefore, the Fermi surface of  finite $t_z$ prefers SDW stripe along diagonal direction. In either case, we emphasize that the period-four SDW arises from an intrinsic strong-coupling mechanism. Perturbations such as $C_4$ symmetry breaking or finite $t_z$ merely select the specific direction of $\mathbf{Q}$, but are not essential for the existence of the SDW itself. In our theory, magnetic order is dominated by local moments in the $d_{z^2}$ orbital, distinct from weak-coupling theory based on  Fermi surface nesting instabilities~\cite{PhysRevB.110.195135,liu2025origin}. We propose that future experiments probing the orbital character of the magnetic moment could verify this mechanism.

\textbf{Conclusion}
In this work, we investigate magnetism and superconductivity in bilayer nickelates using the bilayer type-II $t$-$J$ model, derived from the double Kondo model with strong Hund's coupling $J_H$. Using iDMRG simulations on cylinders with $L_y=2$ and $L_y=4$, we demonstrate the evolution from period-four SDW order to an interlayer paired superconducting phase driven by increasing the interlayer spin coupling $J_\perp$. Our results indicate that local moments from the $d_{z^2}$ orbital are crucial for the formation of magnetic order at small $J_\perp$, implying that a one-orbital model involving only $d_{x^2-y^2}$ fails to capture the SDW physics. Our work establishes the type-II $t$-$J$ model as the minimal framework unifying both the SDW order and superconductivity in bilayer nickelates.

\section{Acknowledgement}
This work was supported by a
startup fund from Johns Hopkins University and the
Alfred P. Sloan Foundation through a Sloan Research
Fellowship (YHZ).
\bibliographystyle{apsrev4-1}
\bibliography{ref}
\clearpage
\onecolumngrid
\appendix
\section{Details of the type-II t-J model}
The localized electron in the $d_{z^2}$-orbital and itinerant electron in the $d_{x^2-y^2}$-orbital can be effectively described within a Kondo-lattice framework. Consequently, the bilayer nickelate system is naturally captured by a double-Kondo model. The double Kondo model reads,
\begin{align}
    H=-t\sum_{\langle ij \rangle;l;\sigma}c^\dagger_{i;l;\sigma}c_{j;l;\sigma}+h.c.+J_\perp\sum_{i}\vec{S}_{i;t}\cdot\vec{S}_{i;b}
    -2J_H\sum_{i;l}\vec{S}_{i;l}\cdot\vec{s}_{i;l;c}+\sum_{\langle ij \rangle;l}J_s\vec{S}_{i;l}\cdot\vec{S}_{j;l}+J_c\vec{s}_{i;l;c}\cdot\vec{s}_{j;l;c},
\end{align}
where $c^\dagger_{i;l;\sigma}$ is the electron operator creates an electron in the $d_{x^2-y^2}$ orbital with spin $\sigma=\uparrow,\downarrow$ in layer $l=t,b$. $\vec{S}$ and $\vec{s}_{c}$ are the spin operators in the $d_{z^2}$ orbital and $d_{x^2-y^2}$ orbital, respectively. $J_\perp$ is the interlayer interaction between the spin moment from the $d_{z^2}$ orbital and $J_H$ is the Hund's coupling between the $d_{z^2}$ and $d_{x^2-y^2}$ orbitals. $J_s$ and $J_c$ are the intralayer interaction between the localized moment in $d_{z^2}$ orbital and itinerant electron $d_{x^2-y^2}$ orbital. In the limit $J_H\rightarrow\infty$, the double Kondo model can be reduced to the bilayer type-II t-J model with spin-$1$ moment $\vec{T}$ and spin-$1/2$ moment $\vec{S}$, as discussed in the main text. The spin-$1$ moment is given by $\vec{T}=\ket{a}\vec{T}_{ab}\bra{b}$, with
\begin{align}
    T_x=\frac{1}{\sqrt{2}}\begin{pmatrix}
        0&1&0\\1&0&1\\0&1&0
    \end{pmatrix},\quad
    T_y=\frac{1}{\sqrt{2}}\begin{pmatrix}
        0&-i&0\\i&0&-i\\0&i&0
    \end{pmatrix},\quad
    T_z=\begin{pmatrix}
        1&0&0\\0&0&0\\0&0&-1
    \end{pmatrix}.
\end{align}

\section{More DMRG results in one-orbital t-J model}
The bilayer one-orbital t-J model with only $d_{x^2-y^2}$ orbital is given by
\begin{align}
    H=-t\sum_{l,\sigma,\langle ij\rangle}c^\dagger_{i;l;\sigma}c_{j;l;\sigma}+J_\parallel\sum_{l,\langle ij\rangle}\vec{S}_{l;i}\cdot\vec{S}_{l;j}+J_\perp\sum_{i}\vec{S}_{i;t}\cdot\vec{S}_{i;b},
\end{align}
where $c^\dagger_{i;l;\sigma}$ is the creation orbital in the $d_{x^2-y^2}$ orbital with spin $\sigma=\uparrow,\downarrow$ at site ${\bf i}$ and layer $l=t,b$ and $\vec{S}_{i;l}$ is the spin operator in $d_{x^2-y^2}$ orbital.

\begin{figure}[h]
    \centering
\includegraphics[width=0.7\linewidth]{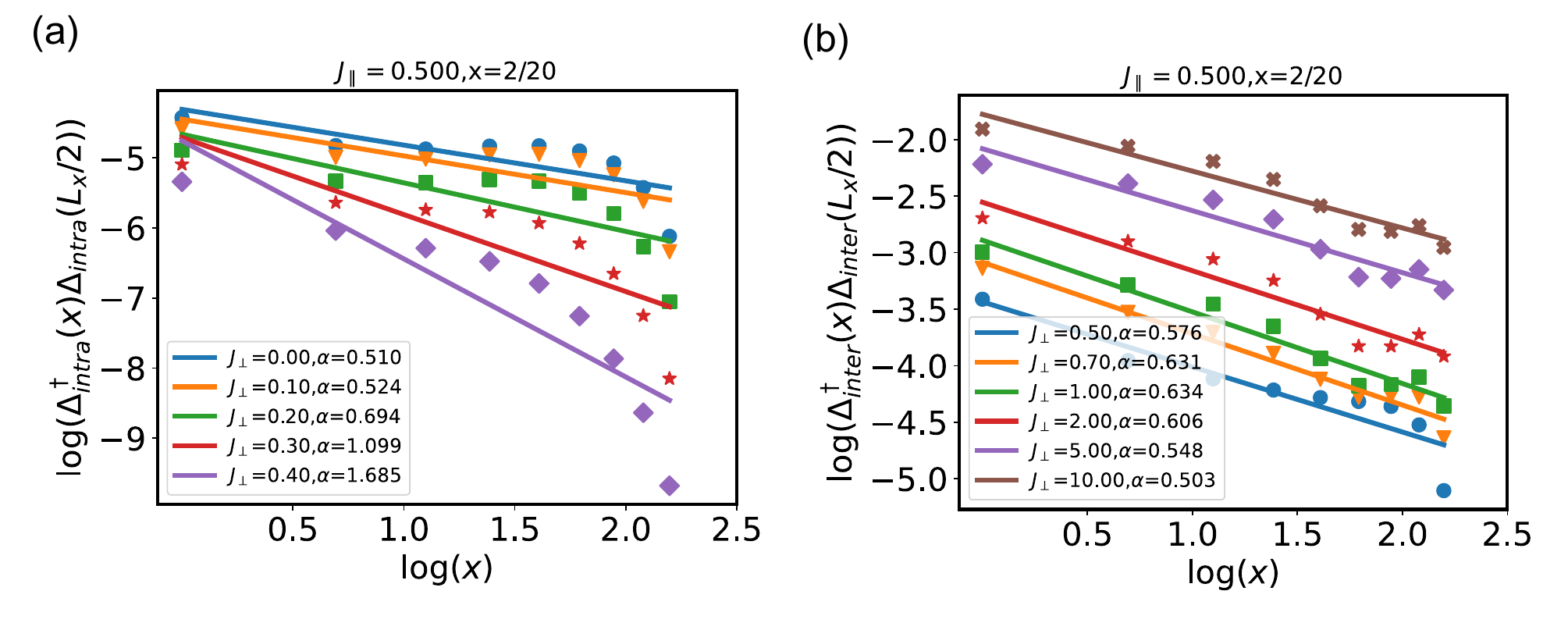}
\caption{Log-log plot of (a) the intralayer and (b) interlayer pair-pair correlation function of one-orbital t-J model for $L_x=20$, $L_y=2$ and $x=0.1$ from finite DMRG with $J_\parallel=0.5$.}
\label{fig:one_orbital_Ly=2}
\end{figure}

The pair-pair correlation function of one-orbital t-J model in $L_x=20$, $L_y=2$ and $x=0.1$ is illustrated in Fig.~\ref{fig:one_orbital_Ly=2}(a) and (b), from which we find, for small $J_\perp\le 0.3$, the system is dominated by the intralayer pairing while for large $J_\perp\ge0.7$, the interlayer pairing dominates.  As shown in Table. I of the main text, the spin gap decreases with $J_\perp$ and then increases again, indicating an intralayer pairing to interlayer pairing transition. The intralayer pairing and interlayer pairing are defined as $\Delta_{intra}(x)=\epsilon_{\sigma\sigma^\prime}c_{y=0,t,\sigma}c_{y=1,t,\sigma^\prime}$ and $\Delta(x)=\epsilon_{\sigma\sigma^\prime}c_{y=0,t,\sigma}c_{y=0,b,\sigma^\prime}$.

\section{Detailed DMRG results for type II t-J model with $L_y=2$ }

\subsection{Detailed DMRG results at small $J_\perp$ and for single layer model}
\begin{figure}[h!]
    \centering
\includegraphics[width=0.8\linewidth]{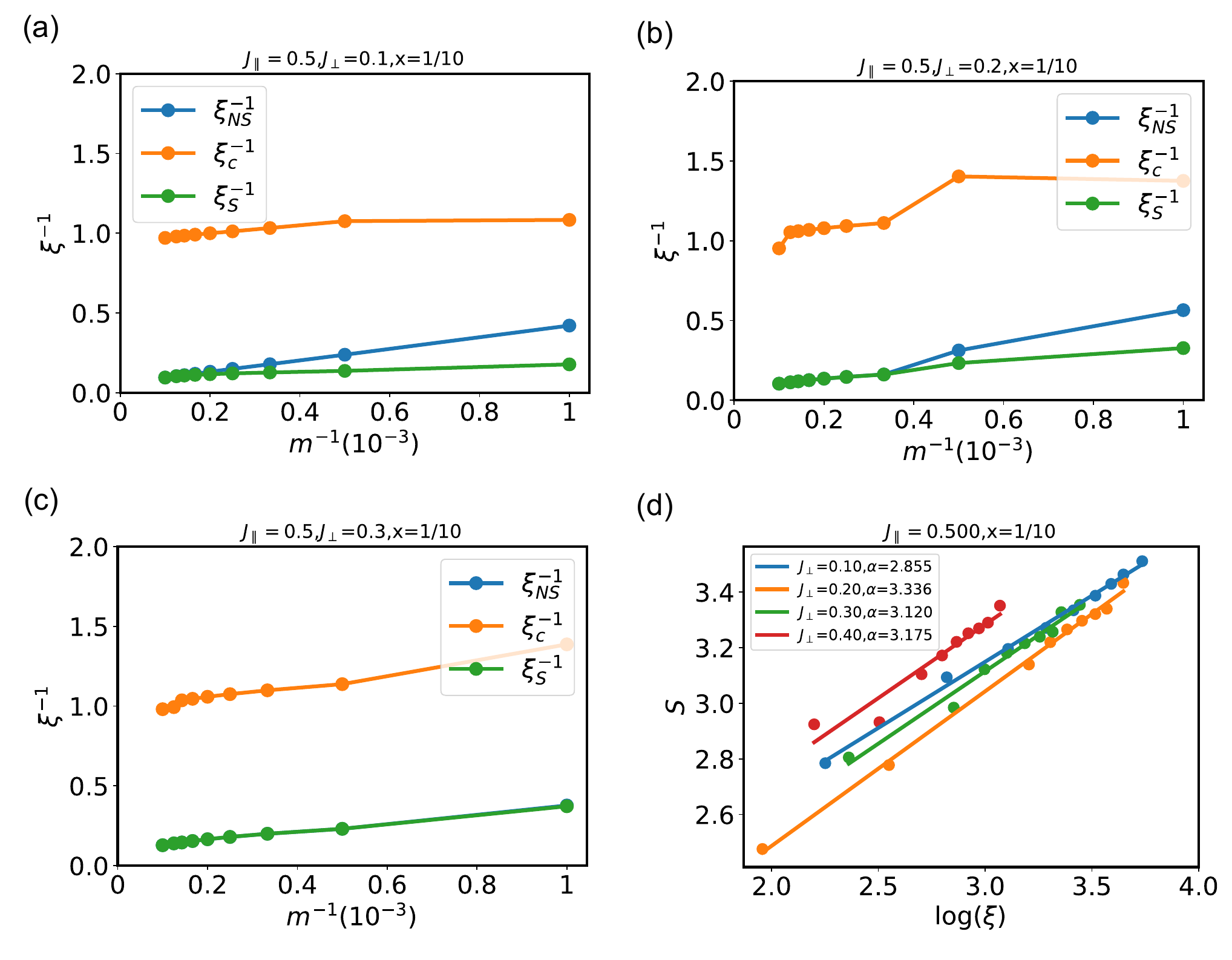}
\caption{(a), (b), (c) correlation lengths of the charge ($N$), spin ($S$), and single-electron ($c^\dagger$) operators from iDMRG simulations of the bilayer type-II t-J model for $L_y=2$ with $J_\parallel=0.5$ at hole doping $x=0.1$. (d) the central charge fitted from $S=\frac{c}{6}\log\xi$. Here the intralayer interaction $J_\parallel^{ss}=J_\parallel^{sd}=0$ and $J^{dd}_{\parallel}=J_\parallel/4.$ }
\label{fig:luttinger_liquid_Ly=2}
\end{figure}

The correlation length of different operators for $x=0.1$ with small $J_\perp\le 0.3$ and $J^{ss}_\parallel=J^{sd}_\parallel=0$ are illustrated in Fig.~\ref{fig:luttinger_liquid_Ly=2}. At $x=0.1$, the spin correlation length grows as the bond dimension is increased at small $J_\perp$, while the correlation length of single electron is finite. We fit the central charge from the relation $S=\frac{c}{6}\log\xi$ as shown in Fig.~\ref{fig:luttinger_liquid_Ly=2}(d), and find the central charge is around $c\approx3$, indicating a multi-component Luttinger liquid phase with three gapless modes.

\begin{figure}[h!]
    \centering
\includegraphics[width=0.9\linewidth]{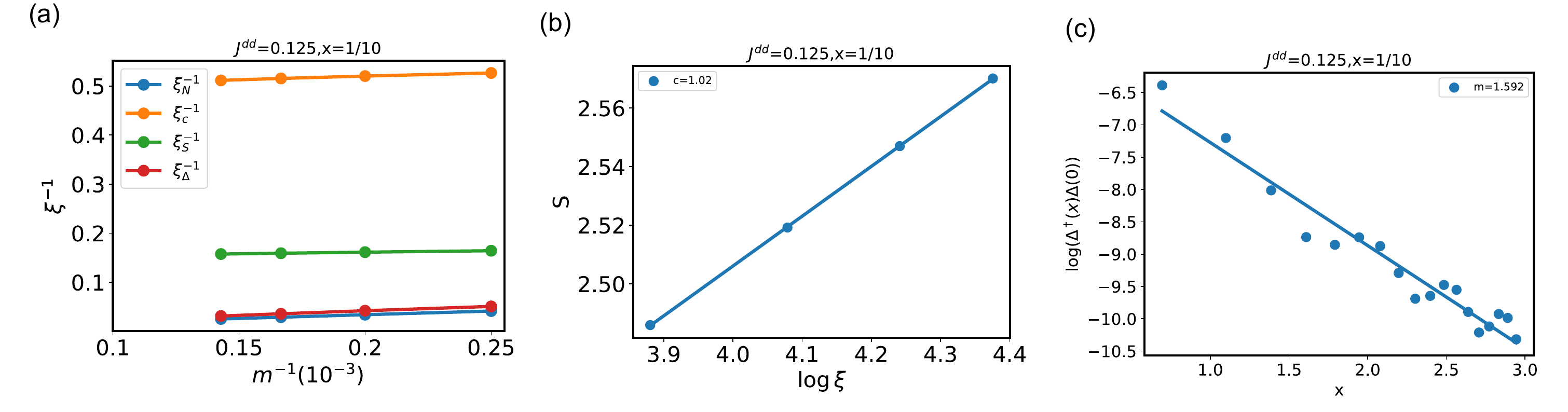}
\caption{ iDMRG result of the single layer type-II t-J model with $J^{ss}_\parallel=J^{sd}_\parallel=0.05$ at hole doping $x=0.1$ in two-leg ladder. (a) the correlation length of different operators, which correspond to different symmetry sectors in the DMRG. (b) central charge $c=1$ fitted from the relation $S=\frac{c}{6}\log\xi$, and (c) log-log plot of the pair-pair correlation function. The pair-pair correlation function exhibits a power-law behavior with exponent $\alpha=1.592$.}
\label{fig:SC_two_leg}
\end{figure}

However, in the presence of a small $J_\parallel^{ss}=J_\parallel^{sd}=0.05$, the single layer model (corresponding to $J_\perp=0$) can be stabilized into the Luther-Emery liquid with intralayer pairing. In Fig.~\ref{fig:SC_two_leg}(a), the correlation length $\xi_{\Delta}$ grows as the bond dimension increases, while the spin correlation length and the single-particle correlation length remain finite.  In Fig.~\ref{fig:SC_two_leg}(b), we fit the central charge with the relation $S=\frac{c}{6}\log\xi$, where $S$ and $\xi$ are the entanglement entropy and correlation length, respectively, and find $c=1$. In Fig.~\ref{fig:SC_two_leg}(c), we fit the pair-pair correlation function and find it exhibits a power-law behavior with exponent $\alpha\approx 1.592$. All these evidences demonstrate the Luther-Emery liquid of the single layer model in two-leg ladder.

\subsection{Detailed results of superconductivity for $L_y=2$ at large $J_\perp$}

\begin{figure}[h]
    \centering
\includegraphics[width=0.7\linewidth]{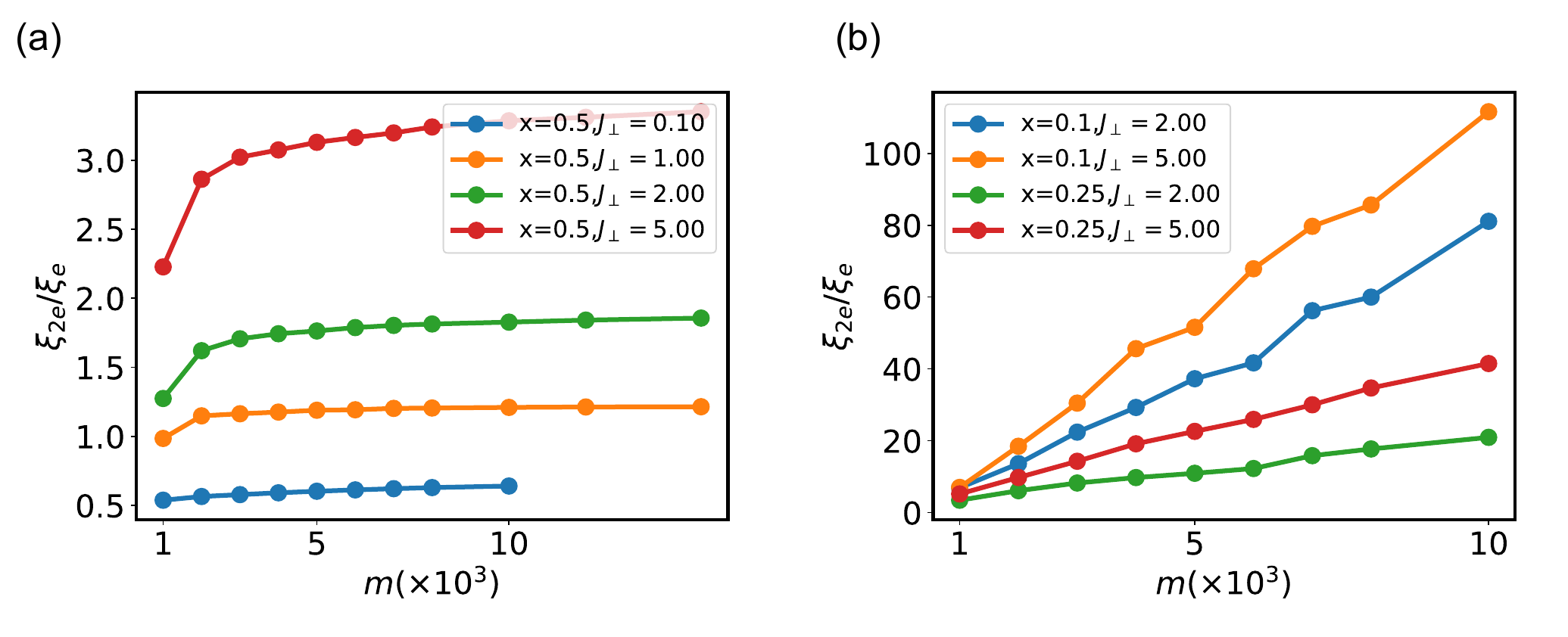}
\caption{The ration $\xi_{2e}/\xi_e$ vs bond dimension for (a) $x=0.5$ and (b) $x=0.1,0.25$ for $L_y=2$. We use the intralayer interaction $J_\parallel=0.5$.}
\label{fig:xi_2e_xi_e_Ly=2}
\end{figure}

In Fig.~\ref{fig:xi_2e_xi_e_Ly=2}, we show the correlation length of $\xi_{\Delta}/\xi_{c}=\xi_{2e}/\xi_e$ for $L_y=2$, where $\xi_\Delta$, and $\xi_c$, are the correlation length of Cooper pair and single electron operators, respectively. We find for $x=0.1$ and $x=0.25$, as we increase the bond dimension, the ratio increases, indicating the system is dominated by Cooper pairs. However, for $x=0.5$, this ratio remains finite, indicating a possible insulating state, which is consistent with the previous study in the ESD model that in two-leg ladder,the ESD model at $x=0.5$ is an insulating state~\cite{oh2025hightemperaturesuperconductivitykineticenergy}.

\section{Detailed  DMRG results for $L_y=4$}
\subsection{Correlation length in the magnetically ordered phase for $L_y=4$}
\begin{figure}[h]
    \centering
\includegraphics[width=0.7\linewidth]{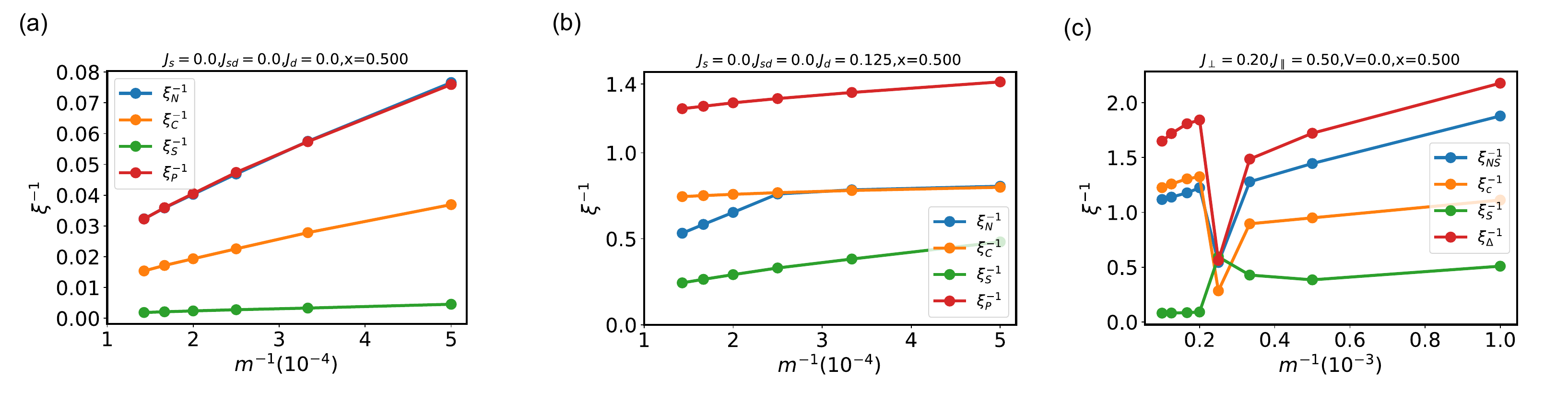}
\caption{The correlation length of different operators in (a) (b) single layer and (c) bilayer type-II t-J model for $L_y=4$ with hole doping $x=0.5$. In (a) we use $J_\parallel=0$ and  in (b) (c), we use $J_\parallel=0.5$. }
\label{fig:corr_len_magnetic_Ly=4}
\end{figure}

Fig.~\ref{fig:corr_len_magnetic_Ly=4} shows the correlation length of different operators of single-layer and bilayer type-II t-J model for $L_y=4$ at $x=0.5$. We can find the spin correlation length increases as the bond dimension increases, indicating the gapless spin degree of freedom. In Fig.~\ref{fig:spin_spin_Ly=4} of the main text, we show that the spin-spin correlation function exhibits a relatively long-range ordered pattern. All these results together demonstrate the existence of magnetically ordered phase in the decoupling or small $J_\perp$ regime.  However, the exact ordering momentum $\mathbf Q$ depends on $J_{\parallel}$ and there is a transition from FM order to SDW with period-four by increasing $J_\parallel$, driven by competition between kinetic-energy-driven FM and $J_\parallel$.
\subsection{More results of interlayer paired superconductivity for $L_y=4$}

\begin{figure}[h!]
    \centering
\includegraphics[width=0.8\linewidth]{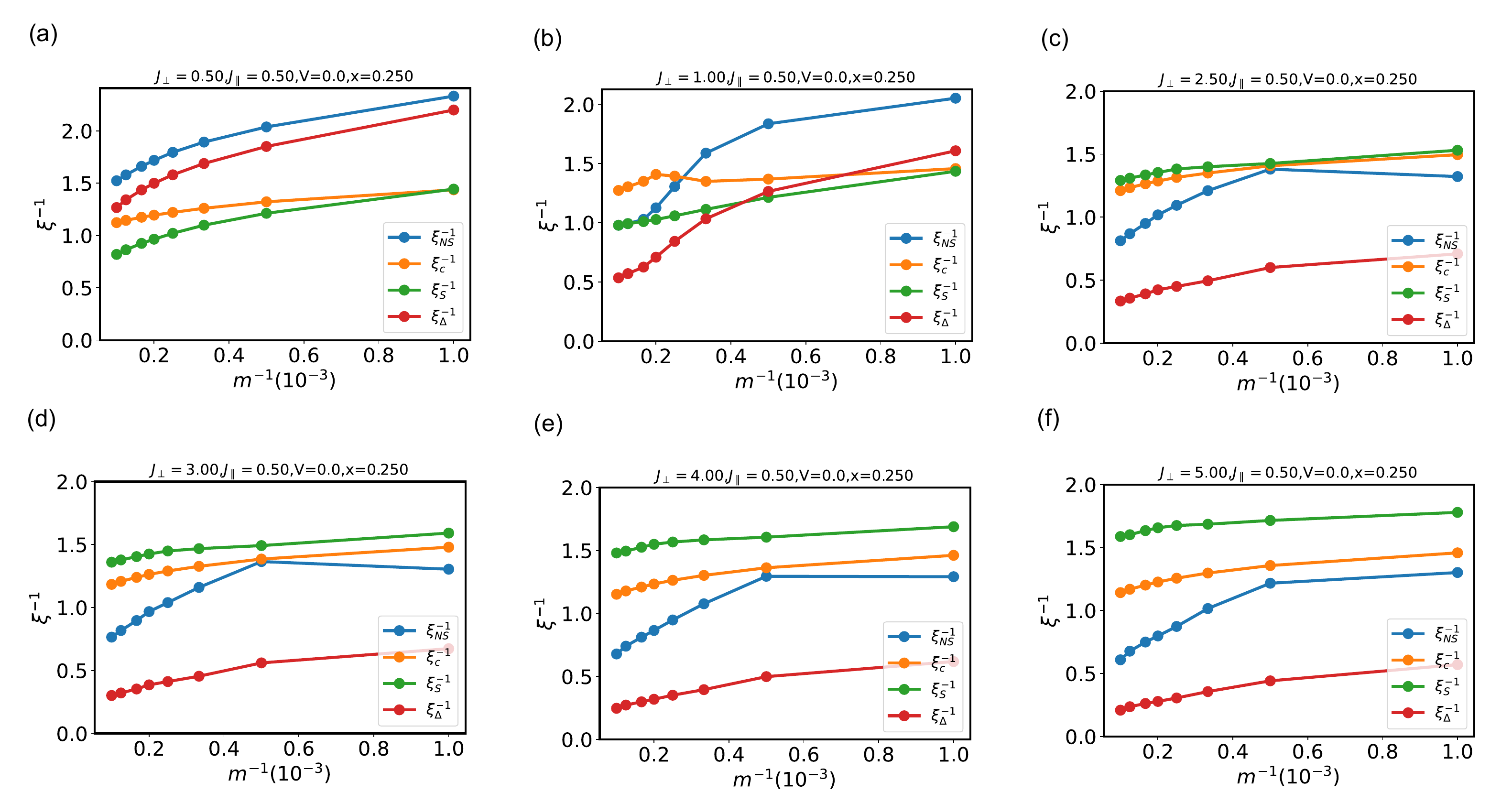}
\caption{The evolution of correlation length of different operators vs bond dimension at $x=0.25$ for $L_y=4$.}
\label{fig:corr_len_Jparallel=0.5_x=0.25_Ly=4}
\end{figure}

\begin{figure}[h!]
    \centering
\includegraphics[width=0.6\linewidth]{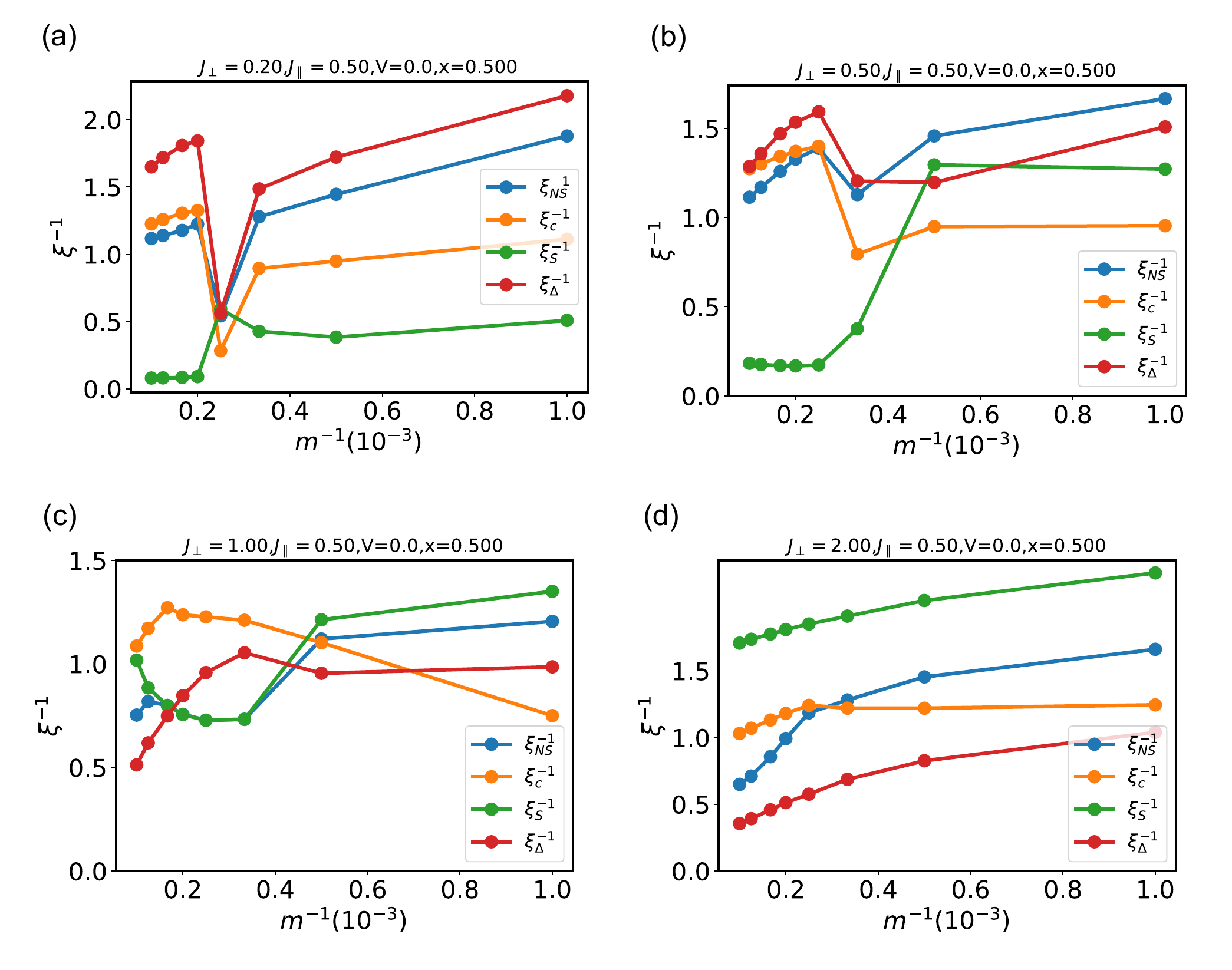}
\caption{The correlation length of different operators vs bond dimension at $x=0.5$ for $L_y=4$. (a)-(d) correspond to $J_\perp=0.2,0.5,1,2$, respectively.}
\label{fig:corr_len_Jparallel=0.5_x=0.5_Ly=4}
\end{figure}

In Fig.~\ref{fig:corr_len_Jparallel=0.5_x=0.25_Ly=4} and Fig.~\ref{fig:corr_len_Jparallel=0.5_x=0.5_Ly=4}, we show the correlation length of different operators for $L_y=4$ with hole doping $x=0.25$ and $x=0.5$, respectively. At small $J_\perp\le 0.5$, the spin correlation length dominates. As $J_\perp$ increases, the correlation length of Cooper pair becomes dominant, indicating the evolution into the superconducting phase when $J_\perp\ge 1$.

\begin{figure}[h!]
    \centering
\includegraphics[width=0.9\linewidth]{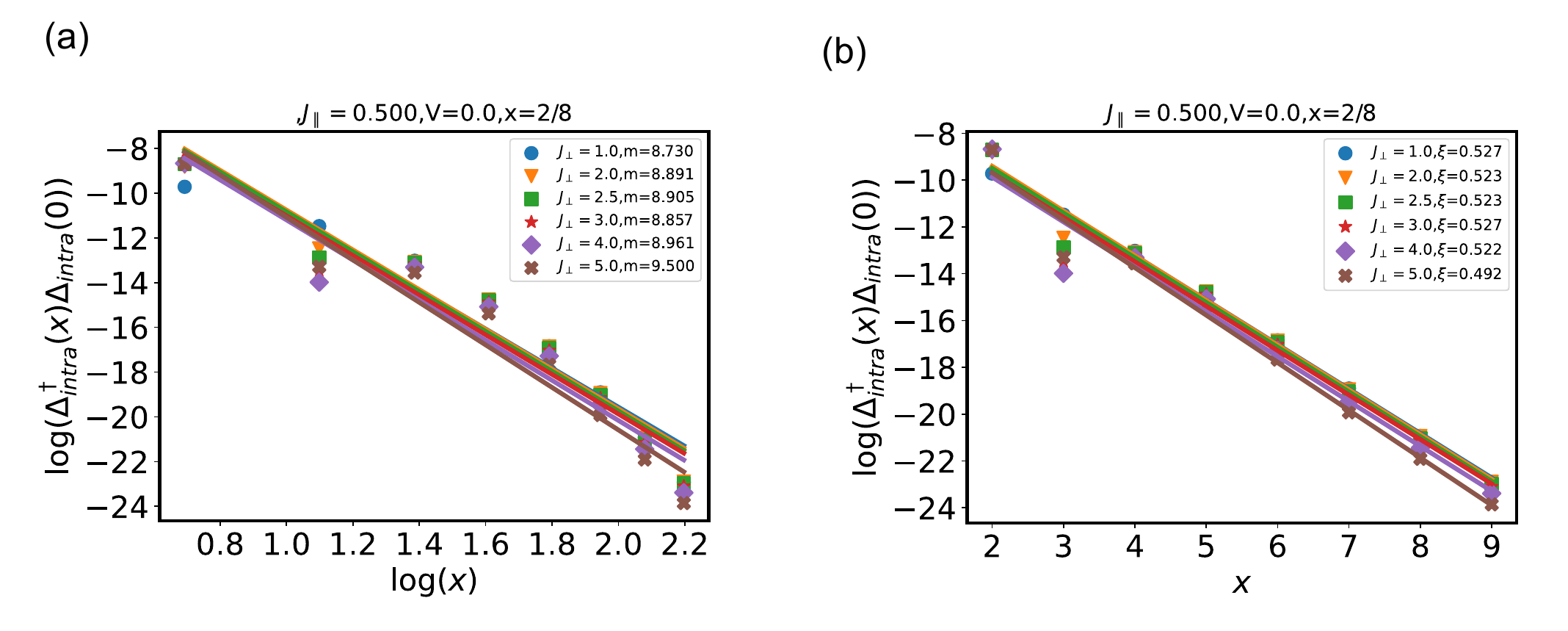}
\caption{(a) the power law and (b) the exponential fit of the intralayer pair-pair correlation function. The intralayer Cooper pair operator is defined as $\Delta_{intra}(x)=\epsilon_{\sigma\sigma^\prime}c_{y=0,t,\sigma}c_{y=1,t,\sigma^\prime}$, where the subscript $y$ corresponds to the coordinate in $y$-direction.}
\label{fig:intralayer_PP_Jparallel=0.5_x=0.25_Ly=4}
\end{figure}

 We still need to check whether the Cooper pair for $J_\perp \geq 1$ is interlayer or intralayer.  In the main text we showed that the interlayer pairing correlation has a power-law decaying with exponent smaller than $2$. Here, we provide the intralayer pair-pair correlation function for $L_y=4$ in Fig.~\ref{fig:intralayer_PP_Jparallel=0.5_x=0.25_Ly=4}. In the plot, we fit the intralayer pair-pair correlation function using both  power-law function (see Fig.~\ref{fig:intralayer_PP_Jparallel=0.5_x=0.25_Ly=4}(a)) and exponential function (see Fig.~\ref{fig:intralayer_PP_Jparallel=0.5_x=0.25_Ly=4}(b)), respectively. We can see that the intralayer pair-pair correlation function is best fit by an exponential decay with a short correlation length $\xi\sim 0.5$.  

\end{document}